# Handover Control for WCDMA Femtocell Networks

**Mostafa Zaman Chowdhury**[*] *Associate Member,*
**Yeong Min Jang**[*] *Lifelong Member*

ABSTRACT

The ability to seamlessly switch between the macro networks and femtocell networks is a key driver for femtocell network deployment. The handover procedures for the integrated femtocell/macrocell networks differ from the existing handovers. Some modifications of existing network and protocol architecture for the integration of femtocell networks with the existing macrocell networks are also essential. These modifications change the signal flow for handover procedures due to different 2-tier cell (macrocell and femtocell) environment. The handover between two networks should be performed with minimum signaling. A frequent and unnecessary handover is another problem for hierarchical femtocell/macrocell network environment that must be minimized. This work studies the details mobility management schemes for small and medium scale femtocell network deployment. To do that, firstly we present two different network architectures for small scale and medium scale WCDMA femtocell deployment. The details handover call flow for these two network architectures and CAC scheme to minimize the unnecessary handovers are proposed for the integrated femtocell/macrocell networks. The numerical analysis for the proposed M/M/N/N queuing scheme and the simulation results of the proposed CAC scheme demonstrate the handover call control performances for femtocell environment.

## 1. Introduction

The upcoming next generation wireless networks are promising to provide high demand of bandwidth with assured quality of service (QoS) for not only voice but also for different data and multimedia services. The tremendously increasing high demand of data rate for wireless communication will be provided by existing and newly developed heterogeneous networks. Femtocell networks have the capability to provide sufficient services at the home environment with lower cost. The closer transmitter and receiver increase the capacity of wireless link and create dual benefits of higher quality links and more spatial reuse [1]. So, femtocell is one of the best approaches for the heterogeneous convergence networks of IMT-Advanced networks.

The network management protocol, network management entity, and network connectivity for femtocell networks differ from the existing networks. Small, medium, and large scale femtocell deployments are possible to reduce the system installation cost. In the small scale deployment, only few numbers of users use the femtocell networks within a macrocell coverage area. Thus, for these small number of femtocell users, large modification in the system is not economically feasible. When the number of user is increased, the system architectures must be modified to manage large amount of femto access points (FAPs). If there are much more overlapping of femtocell coverage, then the management system will be much more complex. This large number of dense and overlapping femtocells can be termed as large and dense deployment. Large deployment without much more overlapping femtocell coverage areas can be termed as medium scale deployment. Thus, different femtocell network architectures will be found in different areas depending on the dense of population, number of internet users, existing network architecture and probability of future extension. The deployable WCDMA femtocell network architecture only considers small scale and medium scale. To support the existing WCDMA networks, we must develop efficient handover system for deployable WCDMA femtocell networks.

Huge number of possible candidate femtocells for macrocell to femtocell handover requires a large neighbor FAP list and communication with many neighbor FAPs for the pre-handover procedure. The optimal solution of this problem can improve the performance of femtocell networks. For the handover procedure, the exchange of information about neighborhood cells and user equipment (UE), pre-authentication, and security confirmation are needed in advance to make seamless and fast handover. As less number of scanning and signaling flows as possible should happen to reduce the power consumption as well as to make the handover fast.

In femtocell/macrocell overlaid networks [15], blocking a macrocell to femtocell handover call does not end the call. The call still connected with macro base station (BS). So, huge bandwidth should not be sacrificed to reduce little handover call blocking

This research was supported by the MKE (Ministry of Knowledge and Economy), Korea, under the ITRC (Information Technology Research Center) support program supervised by the IITA (Institute of Information Technology Assessment) (IITA-2009-C1090-0902-0019).
Kookmin University, Korea, {mzceee, yjang}@kookmin.ac.kr

probability. The optimal number of guard channels can be found and optimization among new call blocking probability, handover call blocking probability, and bandwidth utilization is possible using *M/M/N/N* queuing analysis. The femtocell coverage is small and users always move around the femtocell coverage area. It creates some unnecessary handovers that is a serious problem for femtocell network deployment. These unnecessary handovers cause the reduction of user's QoS level and system capacity. These unnecessary handovers can be minimized using proper call admission control (CAC) and resource management.

The remainder of this paper is organized as follows. In Section 2 we provide the femtocell network architecture for the small and medium scale femtocell network deployment. Call flow for handovers between macrocell and femtocell are presented in Section 3. Section 4 consists the performance analysis of handover call control. The *M/M/N/N* queuing scheme for femtocell environment with numerical results and a CAC to minimize the unnecessary handovers with simulation results are also presented in this section. Finally, we give our conclusion in Section 5.

## 2. System Architecture

The device to core network (CN) connectivity is one of the main concerns about the femtocell network architecture [1]-[3], [11]-[15]. Proper design of this connectivity can solve the major problems regarding security and QoS provisioning issues [11]. Also, the femtocell network architecture may use the access control mechanism [13], [14] to prevent some of the users. Only pre-registered users can access that FAP. For the femtocell/macrocell network integration, several options are possible. Each option comes with a tradeoff in terms of scale but the best option depends on an operator's existing network capabilities and their future plan regarding the network expansion. A femtocell management system (FMS) is used to control and manage the FAPs within an area [2]. The small scale, medium scale or large and dense scale femtocell deployment network architectures differ in terms of network entity, connecting procedures and management systems.

### 2.1 Architecture for Small Scale Deployment

Fig. 1 shows the deployable cost effective small scale WCDMA femtocell network architecture. This architecture is quite similar to the existing 3G network architecture. Each FAP in this architecture is considered as an equivalent of NodeB. Network security can be handled by the IP security protocol between the FAP and the security gateway (SeGW). A femtocell information server (FIS) is connected with the RNC. This FIS stores all information related to the connected FAPs.

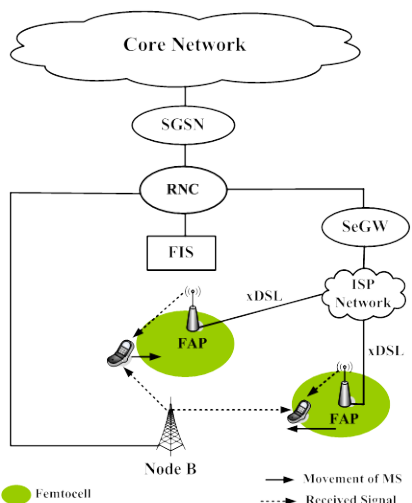

Fig. 1. UE to CN connectivity for fast or small scale integration of the femtocell into existing WCDMA network infrastructures

This architecture is suitable for an operator who has an existing 3G infrastructure deployed; the number of FAPs within the macrocell is not much more; and who is looking for fast integration of the femtocell with the existing infrastructures. This architecture cannot support large number of FAP, because broadcasting such large information through RNC incurs too much overhead. Hence, the handover call flow for small scale femtocell deployment is almost same as existing WCDMA networks.

### 2.2 Architecture for Medium Scale Deployment

Whenever the number of FAP increases in an area, the network architecture, management system is also changed. The traditional WCDMA networks utilize centralized devices, RNCs, to control their associated BSs. One RNC is in charge of radio resource management (RRM) of about 100 BSs [3]. It's not possible to handle or control so many FAPs using the current network control entities. Hence, for medium and large scale femtocell deployment, FAP connectivity should be different than that of existing macrocellular network connectivity. Fig. 2 shows device to CN connectivity for medium and partially large scale femtocell network deployment. This architecture can support large number of FAPs. The femto gateway (FGW) and FMS are the new entities here. Several FAPs are connected to FGW through broadband ISP network. There is no direct connection between the FGW and RNC. They communicate with each other through CN. The FGW can manage thousand of FAPs. Traffic from different FAPs come to FGW and then send to desired RNC, and traffic come from RNC send to target FAP. However, for very dense femtocells, there must need some more complex management and self organized networks (SON) entities.

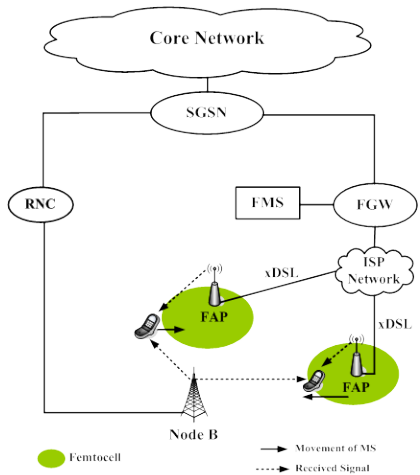

Fig. 2. Femtocell network architecture for medium and partially large scale WCDMA femtocell network deployment

## 3. Handover Call Flow

The ability to seamlessly switch between the femtocell and the macrocell networks is a key driver for femtocell network deployment. However, until now there is no effective and complete handover scheme for the femtocell network deployment. The handover procedures for existing 3GPP networks are presented in [4]-[10]. This section proposes the complete handover call flows for the small and medium scale deployable integrated femtocell/macrocell network architecture. The proposed handover schemes optimize the selection/reselection/RRC management functionalities in the femtocell/macrocell handover. During the information gathering phase, the UE collects information about the handover candidates, and authentications are acquired for security purposes. In handover decision phase, the best handover candidate is determined. Finally, after deciding to perform the actual handover, the UE initiates handover. For the handover between macrocell and femtocell, initial network discovery for femtocell and initial access information gathering are needed. FAP has the RRC functionalities whereas, NodeB has no RRC functionalities. So, the proposed handover call flow for femtocell/macrocell integrated networks differs from that of existing WCDMA macrocellular networks.

An effective call flow sequence within minimum number of signaling is needed for a better mobility management system. For the small scale and medium scale femtocell deployment, macrocell to femtocell and femtocell to macrocell handovers are considered only. Macrocell to femtocell handover is more complex than the femtocell to macrocell handover. Finding the neighboring FAPs and determining the appropriate FAP for handover are challenging for optimum handover decision in macrocell to femtocell handover. In this handover, mobile station (MS) needs to select the appropriate target FAP among many FAPs. Also interference level is considered for handover decision.

### 3.1 Handover for Small Scale WCDMA Femtocell Deployment

The handover for this architecture is simpler because there is not much modification of existing networks system. Also, there is not many target FAPs for macrocell to femtocell handover. FIS is used to provide necessary information during the handover. FIS stores the information about the FAP's identification, registered user's identification, frequency that used by the FAP, and the location of the FAP.

*3.1.1 Handover from Macrocell to Femtocell*

Fig. 3 shows the detail call flow procedures for the macrocell to femtocell handover. Whenever the MS in the macrocell network detects a signal from FAP, it sends a measurement report to the connected NodeB (steps 1, 2). Based on the measurement report, MS decides for handover (step 3), and the NodeB starts handover procedures by sending a handover request to the serving RNC (step 4). The RNC checks the FAP's and user's information from the FIS (steps 5, 6). The handover request is forwarded from the source NodeB to target FAP through CN (Steps 7, 8, and 9). The CAC and RRC are performed only by FAP to check whether the call can be accepted or not (step 10). Then the FAP responses for the handover request (step 11). Steps 12, 13, and 14 are used to setup a new link between RNC and the target FAP. Then the packet data are forwarded to target FAP (step 15). Now the UE re-establishes a channel with the target FAP and detach from the source NodeB, and also synchronized with the target FAP (steps 16, 17, 18, 19, and 20.

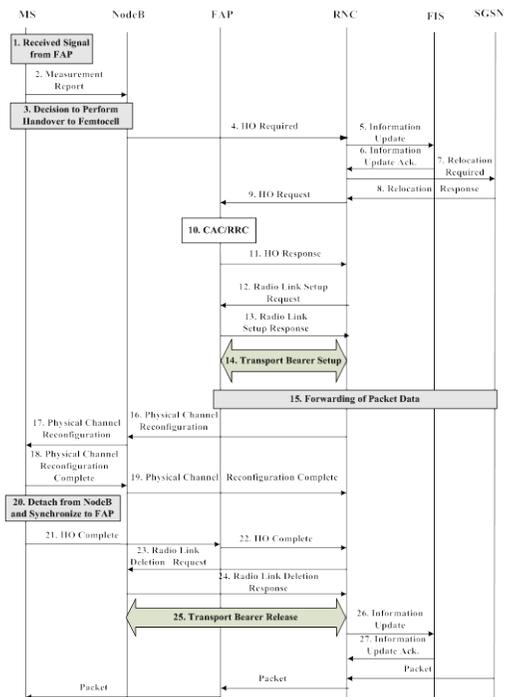

Fig. 3. Call flow for the macrocell to femtocell handover in small scale WCDMA femtocell deployment

MS sends a handover complete message to RNC by informing that, the MS already completed handover and synchronized with the target FAP (steps 21, 22). Then the source NodeB deletes the old link with the RNC (steps 23, 24, and 25). After completing all the procedures, the information in FIS is changed (step 26, 27). Now the packets are sent to MS through the FAP.

*3.1.2 Handover from Femtocell to Macrocell*

Fig. 4 shows the detail handover call flows for femtocell to macrocell handover. If femtocell user detects that femto signal is going down, MS send this report to connected FAP (steps 1, 2). After deciding for handover (step 3), FAP starts handover procedures by sending a handover request to serving RNC (step 4). Steps 5, 6, and 7 show the forwarding the handover request to target NodeB through the CN. The CAC and RRC are performed by the NodeB and RNC to check whether the call can be accepted or not (step 8). Then the NodeB responses for the handover request (step 9). Steps 10, 11, and 12 are used to setup a new link between RNC and the target NodeB. Then the packet data are forwarded to target NodeB (step 13). Now the MS needs to re-establish a channel with the target NodeB and detach from the source FAP, and also synchronized with the target NodeB (steps 14, 15, 16, 17, and 18). MS sends a handover complete message to RNC (steps 19, 20). Then the FAP deletes the old link with the RNC (steps 21, 22, and 23). After completing all the procedures, the information in FIS is changed (step 24, 25). Now the packets are sent to MS through the NodeB.

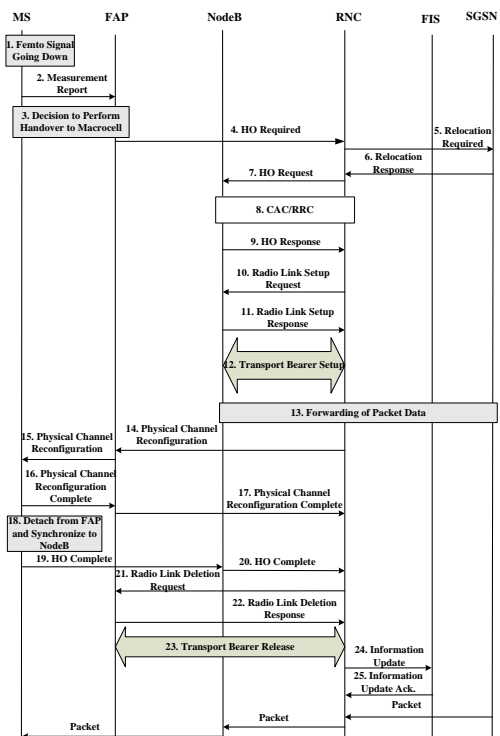

Fig. 4. Call flow for the femtocell to macrocell handover in small scale WCDMA femtocell deployment

## 3. 2 Handover for Medium Scale WCDMA Femtocell Deployment

The handover for this architecture contains more signaling than that of small scale deployable network architecture. Interference level is also considered for this handover. FGW has a strong playing role for this architecture. Appropriate location information is also important for the handover. The message exchange between the FGW and RNC occurs through CN. Each NodeB contains one DB server that stores the neighbor FAP's and registered user's information. This DB server provides exact neighbor FAP list during the handover.

*3.2.1 Macrocell to Femtocell Handover*

Macrocell to femtocell handover is the most challenging issue for medium and large scale femtocell network. In this handover MS needs to select the appropriate target FAP among many candidate FAPs. Also, interference level should be considered for handover decision. Serving NodeB coordinates the handover of MS from NodeB to a FAP by providing information of allowed FAPs to scan for making a FAP neighbor list. Whenever the MS sends the measurement report to FAP, it should also contain the interference level information. The authorization should be checked during the handover preparation phase. Fig. 5 shows the detail call flow procedures for macrocell to femtocell handover in medium scale WCDMA femtocell network. Whenever the MS in the macrocell network detects a signal from femtocell, it sends a measurement report to the connected NodeB (steps 1, 2). Based on the report, MS decides for handover (step 3). The NodeB provides the optimized and authorized neighbor FAP list (step 4). The NodeB starts handover procedures by sending a handover request to the serving RNC (step 5). The handover request is forwarded from the source NodeB to target FAP through the CN and FGW (steps 6, 7, and 8). The FAP checks the user's authorization (steps 9, 10). The FAP performs CAC, RRC and also compare the interference level in current and target femtocell area to admit a call (steps 11). Then the FAP responses for the handover request (step 12, 13, and 14). A new link is established between the FGW and the target FAP (steps 15, 16, 17, 18, and 19). Then the packet data are forwarded to target FAP (step 20). Now the MS re-established a channel with the target FAP, detached from the source NodeB, and synchronized with the target FAP (steps 21, 22, 23, 24, and 25). Then the source NodeB deletes the old link with the RNC (steps 29, 30, and 31). Now the packets are forwarded to MS through the FAP.

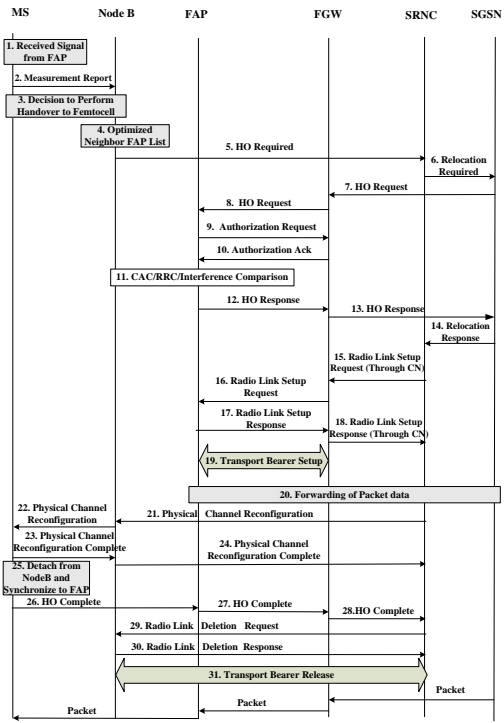
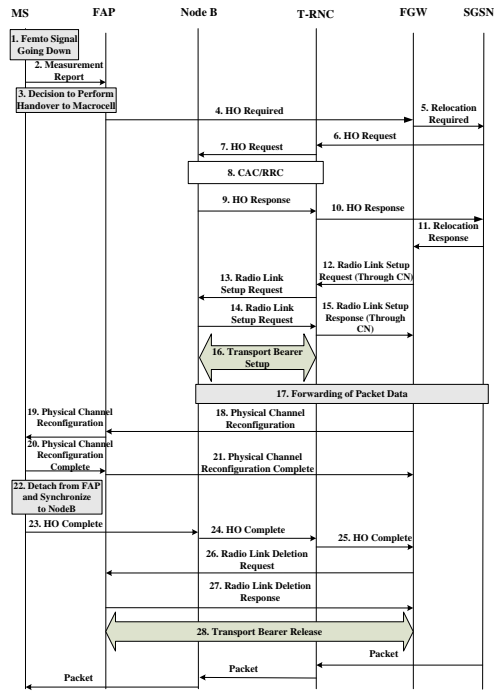

Fig. 5. Call flow for the macrocell to femtocell handover (intra SGSN) in medium scale WCDMA femtocell deployment

*3.2.2 Femtocell to Macrocell Handover*

The handover from femtocell to macrocell is not complex like macrocell to femtocell handover. Fig. 6 shows the detail handover call flow procedures for femtocell to macrocell handover in medium scale deployment. If femto user detects that femto signal is going down, MS send this report to connected FAP (steps 1, 2). After deciding (step 3) for handover, FAP starts handover procedures by sending a handover request to serving RNC (step 4). Steps 5, 6, and 7 show the forwarding the handover request to target NodeB through the CN. The CAC and RRC are performed to check whether the call can be accepted or not (step 8). Then the NodeB responses for the handover request (step 9). Steps 10, 11, 12, 13, 14, 15, and 16 are used to setup a new link between RNC and the target NodeB. The packet data are forwarded to target NodeB (step 17). The MS re-establishes a channel with the target NodeB and detached from the source FAP, and also synchronized with the target NodeB (steps 18, 19, 20, 21, and 22). MS sends a handover complete message to RNC to inform that, the MS already completed handover and synchronized with the target NodeB (steps 23, 24, and 25). Then the FAP deletes the old link with the RNC (steps 26, 27, and 28). Now the packets are sent to UE through the NodeB.

Fig. 6. Call flow for the femtocell to macrocell handover (intra SGSN) in medium scale WCDMA femtocell deployment

The proposed handover call flows follow all the basic requirements of a successful handover [4]-[10] to make a seamless and reliable handover. The proposed schemes consider signal-to-interference (CIR) level, and users' authentication that reduces the number of target FAPs for the handover candidate. This causes the lowest number of neighbor list and reduction of signaling overhead. Some unnecessary signaling in existing handover schemes [4]-[10] are not considered to reduce the signaling overhead. Hence, our proposed handover call flow schemes provide reliable handover with minimum signaling overhead.

## 4. Performance Analysis of Handover Control

The call arriving rate in different femtocell environment is different. Thus the handover call management differs in different femtocell environment. In femtocell/macrocell overlaid networks, blocking a macrocell to femtocell handover call is not dropping that call. By reserving large amount of bandwidth for the macrocell to femtocell handover calls, huge bandwidth utilization should not be sacrificed to reduce little handover call blocking probability. Frequent and unnecessary handovers causes the reduction of QoS. The efficient handover scheme to increase the bandwidth utilization by optimizing handover call blocking probability, and CAC scheme to reduce the unnecessary handovers are proposed in this section.

### 4.1 Calculating Handover Call Blocking Probability

Fig. 7 shows the basic state transition rate diagram for femtocell layer. The femtocell layer is proposed here by

*M/M/N/N* queuing system [12]. In the Fig. 7, $\lambda_{hm}$, $\lambda_{nf}$, and $\mu$ represents the average rate of handover call from macrocell to femtocell, average originating new calls rate at femtocell area, and service rate respectively. A system with the value of *K* less than *N* implies that the system gives more priority to macrocell to femtocell handover calls than the originating new calls at femtocell area. The fixed value of *K* for all femtocell environments reduces the resource utilization. Several schemes or techniques may be taken to fix the value of *K*. Suppose for a femtocell coverage area with less probability of handover call rate, can use very close value of *K* and *N*. The system can also use a variable value of *K* to optimize between the resource utilization and handover call blocking probability. Thus, the total arrival rate of the connection request of the system is

$$\lambda_j = \begin{cases} \lambda_{nf} + \lambda_{hm} & \text{for } 0 \le j < K \\ \lambda_{hm} & \text{for } k \le j < N \end{cases} \quad (1)$$

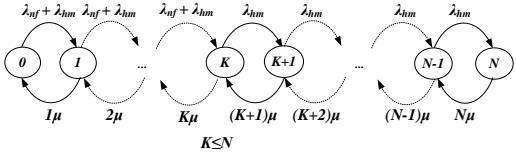

Fig. 7. State transition rate diagram for femtocell layer

The blocking probability of the calls originating at femtocell area is

$$P_B = \sum_{i=K}^{N} \frac{(\lambda_{nf} + \lambda_{hm})^K \lambda_{hm}^{i-K}}{i!\,\mu^i} P(0) \quad (2)$$

The blocking probability of the macrocell to femtocell handover calls is

$$P_D = \frac{(\lambda_{nf} + \lambda_{hm})^K \lambda_{hm}^{N-K}}{N!\,\mu^i} P(0) \quad (3)$$

where

$$P(0) = \left[ \sum_{i=0}^{K} \frac{(\lambda_{nf} + \lambda_{hm})^i}{i!\,\mu^i} + \sum_{i=K+1}^{N} \frac{(\lambda_{nf} + \lambda_{hm})^K \lambda_{hm}^{i-K}}{i!\,\mu^i} \right]^{-1}$$

The priority of the macrocell to femtocell handover calls should not be compromised with sacrificing huge resources. We may find the optimal value of *K* using *M/M/N/N* queuing analysis. Through analysis, we may optimize among new call blocking probability, handover call blocking probability, and bandwidth utilization.

We performed the numerical analysis to find the optimal value of *K*. We assume, the maximum number of users with the FAP is 10, the average service time at femtocell coverage area of each call is 120 sec. Fig. 8 shows the numerical results for the *M/M/N/N* scheme in femtocell environment. For $\lambda_{nf}=0.1$ and $\lambda_{hm}=0.075$, an optimal value of *K* is found to be 8 that is shown in Fig. 8(a). In this case, by sacrificing 2.5% of resources, we can reduce 53% handover call blocking probability. Fig. 8(b) shows that, guard channels are not required for $\lambda_{nf}=0.03$, $\lambda_{hm}=0.01$. In this case, handover call blocking probability is acceptable even there is no guard channel. The guard channels only increase the new call blocking rate and reduce the bandwidth utilization in this case

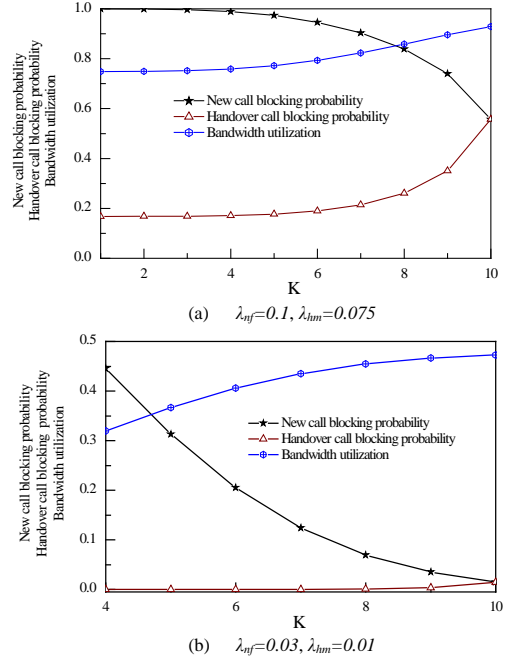

Fig. 8. Optimizing among new call blocking probability, handover call blocking probability, and bandwidth utilization to select the preferable value of K for guard channel

### 4.2 Minimization of Unnecessary Handover

The mobile users always move around the femtocell coverage area. These matters cause some unnecessary handovers in femtocell/macrocell integrated networks. Frequent and unnecessary handover is serious problem for femtocell networks environment, as femtocell coverage area is very small and there is possibility to stay very small time whenever a high speed MS enters into femtocell coverage area. A high speed MS causes two unnecessary handovers due to movement from macrocell to femtocell and again femtocell to macrocell. In the wireless communication systems, the frequent and unnecessary handovers reduce the end-to-end QoS level as well as decrease the capacity of the system. So, the minimization of unnecessary handovers is absolutely necessary for the integrated femtocell/macrocell network system. Whenever a MS is connected with macrocellular network, the MS found change of signal level from FAP due to the movement of MS. Sometimes MS with higher velocity causes very little time to stay in a femtocell coverage area. This causes unnecessary handovers that is indicated by "A" in Fig. 9. In Fig. 9 "B" indicates the case when a MS just move inside the femtocell coverage area and maintain good received signal level for long time. The "C" shown in Fig. 9 indicates the case when a MS moves to femtocell area but does not enter into center area and stay at the boundary area for long time. Hence different types of conditions arise. Due to arising of different conditions, only a unique handover decision making policy is not sufficient to improve the performance. We proposed a CAC scheme to reduce the unnecessary handovers.

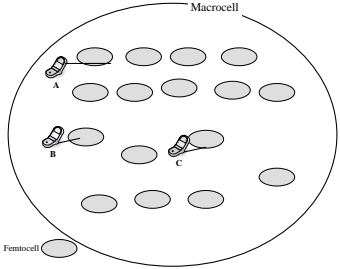

Fig. 9. Movements of MS within macrocell/femtocell coverage area

The number of detected handovers *(h)* in a femtocell coverage area is a function of femtocell radius *(r)*, speed *(v)* of the MS, and angle of movement *(θ)* with respect to the direction of FAP. Thus, the number of handover can be written as

$$h = f(r, v, \theta) \quad (4)$$

It can also be expressed as

$$h \propto \frac{v \sin \theta}{r} \quad (5)$$

A proper CAC can reduce the number of unnecessary handovers within a femtocell/macrocell integrated network. Fig. 10 shows the proposed CAC to reduce the number of unnecessary handovers whenever a macrocell user moves to femtocell coverage area. The decision of handover can be taken using the decision parameter $X$

$$X = S_{femto} V_{mobile} CIR_{femto} \quad (6)$$

In (6), $S_{femto}$ represents the received signal strength indicator *(RSSI)* from *FAP*, $V_{mobile}$ represents the velocity of MS, and $CIR_{femto}$ represents the CIR at femtocell area. The value of $S_{femto}$ is 1 only if the received signal level does not go below a threshold level for a specific time interval, else it is 0. $V_{mobile}$ represent 1 if velocity of MS is less than a threshold velocity, otherwise 0. The value of $CIR_{femto}$ is 1 either the CIR at femtocell environment is greater than the threshold value or CIR at macrocell environment, else its value is 0. Thus the value of decision parameter is either 0 or 1. A macrocell to femtocell handover call is only accepted by FAP if the value of $X$ is 1.

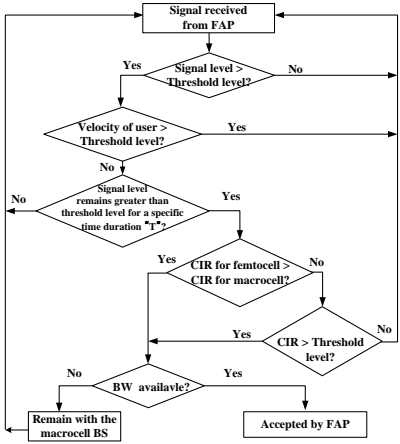

Fig. 10. CAC to accept a handover call by FAP

For the proposed CAC, four parameters such as, received signal level, duration of time a MS maintains the minimum required signal level, velocity of MS, and CIR level are considered. The threshold level of signal is the minimum level of signal that must be needed to handover a MS from macrocell to femtocell. Sometimes MS receives the signal greater than minimum required level but within very short time the level again go down due to the movement of the MS. Whenever a MS moves to femtocell area, the MS must maintain threshold level of signal for minimum threshold "T" time. A call can be accepted if, either CIR level in the target femtocell is less than the threshold level or less than the CIR level of current macrocell area. The threshold time "T" can be chosen according to the service type, QoS requirement, and the velocity of the user. Suppose, data users are delay and bandwidth adaptive. Extra handover will not degrade the QoS level much more for the data user compare to the voice users. Thus, for the data user, small threshold time can be chosen to provide higher throughput. Because, the FAP supports higher data rate for data users even for short duration of time. However, very long threshold time for voice user will not cost effective and will not provide better utilization of femtocell's resources.

We verified the performance of the proposed CAC scheme using simulation result. Table 1 shows the basic simulation parameters. We randomly generate the angle of movement of a MS. The apparent stay time in the femtocell coverage area of a MS is calculated from the velocity and the movement direction. We consider 150 FAPs within a macrocell coverage area. In our simulation, we consider a handover as an unnecessary handover when the MS move from macrocell to femtocell and within 40 seconds it moves to macrocell again or within 10 seconds it terminates the call.

Table 1. Simulation parameters

| | |
|---|---|
| Radius of femtocell coverage area [m] | 10 |
| Average velocity of MS in femtocell coverage area [km/hr] | 1 |
| Type of service | Voice |
| Average call life time after handover from macrocell to femtocell [sec] | 90 |
| Call life time and user velocity | Exponential distribution |
| User movement direction | Random |
| Number of FAP within a macrocell | 150 |
| Threshold velocity [km/hr] | 10 |

Fig. 11 shows the number of handover from macrocell to femtocell or again femtocell to macrocell for different schemes. This shows that, all the users move from macrocell to femtocell coverage area does not need to handover from macrocell to femtocell. Our proposed scheme optimized many unnecessary handovers. In Fig. 11 a traditional scheme without any threshold time causes much more unnecessary handovers than proposed schemes. Fig. 12 shows the unnecessary handovers minimization of the proposed scheme. It shows that the CAC that does not consider any threshold time, makes about 38% unnecessary handover. A threshold time of 20 sec and 10 sec reduces the unnecessary handover into 8% and 19% respectively.

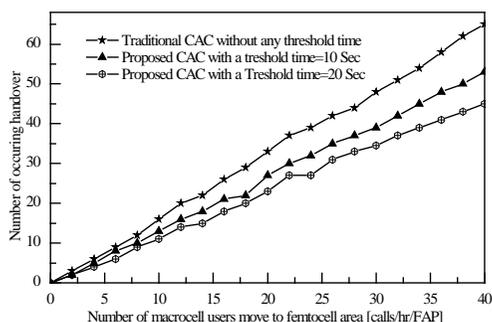
Fig. 11. Observation of the number of occurred handover whenever the users move from macrocell to femtocell coverage area

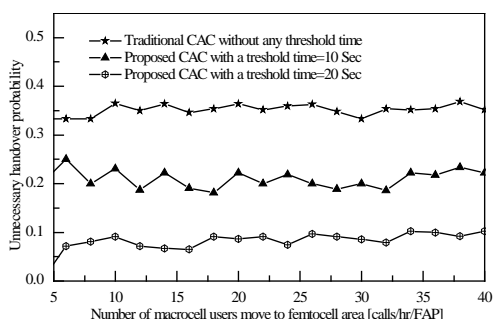
Fig. 12. Observation of occurred unnecessary handover probability whenever the users move from macrocell to femtocell

## 5. Conclusion

The integrated femtocell/macrocell network is the attractive solution for the future convergence networks. It can provide higher QoS for indoor users at low price, while simultaneously reducing the burden on the whole network system. However, handover call control is one of the challenging issues for the effective deployment of integrated femtocell/macrocell networks. Minimum number of signaling during handover, seamless handover, fast handover, and the unnecessary handover minimization are the main concern for mobility management for this integrated network.

We consider only small scale and medium scale deployment of WCDMA femtocell networks. The proposed handover call flows are explained in details. These handover call flows are able to provide a seamless and reliable handover between macrocell and femtocell both for the small scale and medium scale. The proposed *M/M/N/N* queuing scheme for femtocell environment optimizes among the new call blocking probability, handover call blocking probability, and bandwidth utilization. The proposed CAC is able to reduce the unnecessary handovers. The simulation results showed that the proposed unnecessary handover minimization scheme is an effective scheme to reduce the number of unnecessary handovers.

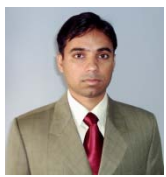

**Mostafa Zaman Chowdhury** received B.Sc. in Electrical and Electronic Engineering from Khulna University of Engineering and Technology (KUET), Bangladesh in 2002. Then, he joined as a faculty member in Electrical and Electronic Engineering department of KUET, Bangladesh in 2003. He completed his MS from Wireless Networks and Communications Lab. of Kookmin University, Korea in 2008. Currently he is continuing his Ph.D. studies in Wireless Networks and Communications Lab. of Kookmin University, Korea. His current research interests focus on convergence networks, QoS provisioning, mobility management, and femtocell networks.

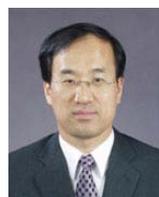

**Yeong Min Jang** received the B.E. and M.E. and M.E. degree in Electronics Engineering from Kyungpook National University, Korea, in 1985 and 1987, respectively. He received the doctoral degree in Computer Science from the University of Massachusetts, USA, in 1999. He worked for ETRI between 1987 and 2000. Since Sept. 2002, he is with the School of Electrical Engineering, Kookmin University, Seoul, Korea. His research interests are IMT-advanced, radio resource management, convergence networks, and femtocell networks.